\documentclass{elsart}
\usepackage[dvips]{graphicx}
\usepackage{amsmath}
\begin{document}

\title{Non-extensive distributions for a relativistic Fermi Gas}
\author{J. Ro\.zynek}
\address{National Centre for Nuclear Research,
        Department of Fundamental Research,\\ Ho\.za 69, 00-681
        Warsaw, Poland; e-mail: Jacek.Rozynek@fuw.edu.pl
                  }

 {\today}

{\scriptsize Abstract: Recently the non-extensive approach has been used in a variety
of ways to describe dense nuclear matter. They differ in the methods of introducing
the appropriate non-extensive single particle distributions inside a relativistic
many-body system, in particular when one has to deal both with particles and
antiparticles, as in the case of quark matter exemplified in the NJL approach. I
present and discuss in detail the physical consequences of the methods used so far,
which should be recognized before any physical conclusions can be reached from the
results presented.


 }


It is nowadays widely accepted that in many branches of physics experimental data
indicate the necessity of a departure from the standard extensive Boltzman-Gibs
statistics, which is then replaced by a non-extensive statistics \cite{T}. In high energy
scattering one uses Tsallis statistics and Tsallis distributions which describe particle
distributions over a large energy region with the help of only one single additional
parameter, the non-extensivity $q\neq1$. This replaces all additional parameters
necessary when the extensive description is used; for $q=1$ one recovers the usual BG
statistics (cf. \cite{qparticle} and references therein). When considering dense Nuclear
Matter (NM) some kind of mean field theory is used, either in terms of nuclear degrees of
freedom (like the Walecka model (WM) \cite{SW}), or in terms of quark and anti-quark
degrees of freedom (like in the Nambu-Jona-Lasino (NJL) model \cite{NJL,NJLadd,Sousa}).
Both descriptions has been reformulated using a non-extensive approach (for the $q$-WM in
\cite{Pereira,SPSA,LPQ,LP} and for the $q$-NJL model in \cite{JG})\footnote{Other
approaches to non-extensive dense matter can be found in \cite{BUS,AD1,CW}.}. All these
models try to find the non-extensive traces in the nuclear Equation of State (EOS) using
the non-extensive single particle distributions. However, it is very difficult to compare
their results in a conclusive way
 because they use different forms of single, non-extensive particle
distributions. In this work, analyzing a Fermi Gas model we compare and discuss these
distributions following the method presented in \cite{TPM}, both for the momentum
dependent fermion distributions $n_q(p)$ and for anti-fermion distributions
$\bar{n}_q({\bar{p}})$.

In the extensive approach to dense, hot matter the particle and antiparticle
occupation numbers, $n_i$ and $\bar n_i$, can be obtained from the Jayne's
extremalization of the entropic measure
\begin{equation}
S =  \sum_{i} \left[ n_{i} \ln n_{i}+(1-n_{i})\ln (1-n_{i})
   \right] + \left[ n_{i}\rightarrow \bar n_{i}
   \right],\label{extentropy}
\end{equation}
under the constraints imposed by the total number of particles, ${N}$, and the total
energy of the system, ${E}$ ($\epsilon_i$ is the energy of the $i$-th energy level)
\cite{EM},
\begin{equation}
\sum_{i} \left(n_i - \bar{n}_i\right) = N\qquad {\rm and}\qquad
\sum_i \left(n_i + \bar{n}_i \right)\epsilon_i = E.
\label{constraints}
\end{equation}
As a result we get the Fermi-Dirac distributions,
\begin{eqnarray}
\qquad n_{i} &=& \frac{1}{ \exp \left( x_i\right) + 1},\qquad \quad \bar n_{i} =
\frac{1}{\exp \left( \bar{x}_i \right) + 1}, \label{FD}
\end{eqnarray}
which depend on the dimensionless quantities
\begin{eqnarray}
x=\beta(\epsilon-\mu)~~~~~{\rm and}~~~~~~\bar{x}=\beta({\epsilon}+\mu).\label{x}
\end{eqnarray}
where $\beta=1/T$, $\epsilon=\sqrt{p^2+m^2}$, $m$ is  fermion mass and $\mu$ its
chemical potential.
For $\epsilon=0$, the distributions $n(x)$ and $\bar{n}(\bar{x})=n(\bar{x})$ satisfy
  following relation:
\begin{equation}
n(x) + {n}(\bar{x})=1.  \label{prop}
\end{equation}
The chemical potential (separation energy) for anti-fermions is negative $-\mu$ (because
the N increases after removing the anti-fermion), therefore this relation allows holes
among the negative energy states to be interpreted as antiparticles with positive energy.
As we shall see below, this is usually not the case for non-extensive distributions
\cite{BSZ}. The similar relation for particle distributions
\begin{equation}
n(x) + n(-x)=1  \label{prop3}
\end{equation}
exhibits, particle-hole symmetry around the Fermi surface where $x=0$.

 When formulating a non-extensive version of the model of dense matter one must do it in
a thermodynamically consistent way. This means that one has to preserve the standard
thermodynamical relationships among the thermodynamical variables such as, for example,
entropy, energy and temperature,
\begin{equation}
\left(\frac{\partial S}{\partial E}\right)_{V,N} =
\frac{1}{T}.\label{thercons}
\end{equation}
The best solution is to use once more Jayne's maximum entropy prescription and
extremalize the appropriate non-extensive entropic measures with  specifically chosen
constraints because, as was shown in \cite{MEnt_cons}, any thermostatistical formalism
constructed by this method complies with the thermodynamical relationships \cite{TPM}. A
thermodynamically consistent formulation is then obtained by the appropriate
identification of relevant constraints with extensive thermodynamical quantities (like
number of particles $N$ or energy $E$) and of the corresponding Lagrange multipliers with
appropriate intensive thermodynamical quantities (like temperature $T$ and chemical
potential $\mu$).
 In order to describe non-extensive many-body systems characterized by $q\neq1$,
 one has to introduce the non-extensive measure of the entropy expressed, in analogy with
 Eq. (\ref{extentropy}),  in terms of single particle distributions
 for fermions and anti-fermions, with accordingly modified constraints (\ref{constraints}).
In this way we obtain a q-generalized quantum distributions from a given form of the
non-extensive entropy.

There are different choices of non-extensive entropy
 which lead to different thermodynamical solutions \cite{PPMU,TPM,consistency}. We
 shall
discuss three possibilities, two of which are currently exploited \cite{LPQ,AD1}. In the
first choice, one uses a straightforward $q$-generalization of the entropic measure used
in Eq.(\ref{extentropy}), which now has the following form:
\begin{eqnarray}
S_q&=&{\sum_i}s_{qi}~~~~~~~{\rm where~~for} \hspace{.5cm} 0\!<\!n_{qi}(\bar{n}_{qi})\!<\!1 \nonumber\\
(\bullet)~~~~{s}_{qi}&=&\frac{1-n_{qi}^q-(1\!-\!n_{qi})^q}{q-1}+
\left\{n_{qi}\rightarrow\bar n_{qi} \right\}. \label{s1}
\end{eqnarray}

This generalization consists in replacing in Eq.(\ref{extentropy}) $\ln(n)$ by
$\ln_q(n_q)=(1-n_q^{1-q})/(1-q)$ and in using as the corresponding effective occupation
numbers of particles and antiparticles, $n^q_{qi}$ and  $\bar{n}^q_{qi}$, instead of
$n_{i}$ and $\bar{n}_{i}$. This reflects the fact that non-extensive properties can be
formally understood as a presence of some effective interaction between the constituents
of the system, the strength of which is proportional to $|q-1|$ (in the non-extensive
environment, parameterized by $q$, constituents are no longer free particles)
\cite{qBiro}. Therefore  constraints (\ref{constraints}) for the total energy $E$ of the
system and the conservation of the total fermion number $N$ will now have the following
form \footnote{~This form assures the fulfillment of basic requirements of
thermodynamical consistency \cite{PPMU,TPM,consistency}.}:
\begin{equation} \sum_i \left( n_{qi}^q - \bar{n}_{qi}^q\right) =
{N}\qquad{\rm and}\qquad \sum_i \left( n_{qi}^q + \bar{n}_{qi}^q\right)\epsilon_i =
{E}. \label{qconstraints}
\end{equation}

 The extremalization of entropy given by Eq.(\ref{s1}), performed under constraints (\ref{qconstraints}), with respect
to two independent variables, $n$ and $\bar n$, gives the following set of equations
with Lagrange multipliers $\alpha$ and $\beta$:
\begin{eqnarray}
\frac{\partial s_q}{\partial n_q}\! &=& \!(\alpha\!+\!\beta \epsilon )q
n_q^{(q-1)};~~\frac{\partial s_q}{\partial \bar n_q}=(-\alpha\!+\!\beta \epsilon )q
\bar n_q^{(q-1)}
 \hspace{0.3cm}{\rm for}~~0\!<\!n_q(\bar{n}_q)\!<\!1, \label{e1}
\end{eqnarray}
Solving these one obtains the following single particle distributions for fermions and
anti-fermions:
\begin{eqnarray}
n_{q}&=&\frac{1}{e_q(x)\!+\!1},\qquad \bar{n}_{q}=\frac{1}{e_q(\bar{x})\!+\!1}\qquad
{\rm with} \quad e_q (x)=[1+(q\!-\!1)x]^{\frac{1}{q\!-\!1}} , \label{qFD}
\end{eqnarray}
where $x$ and $\bar{x}$ are the same as in Eq.(\ref{x}) with chemical potential
$\mu=-\alpha/\beta$. However, these solutions must be supplemented by conditions assuring
that $\e_q(x)$ in Eq.(\ref{qFD}) is properly defined, i.e. that
$$1+(q-1)x\geq0$$ for all $q$ and $x(\bar{x})$. We get therefore four constraints for particle momenta $p$ and
antiparticle momenta $\bar{p}$ with $q<1$ and $q>1$:
\begin{eqnarray}
\sqrt{p^2 + m^2}&\leq&- \frac{T}{q-1} + \mu,\qquad \sqrt{\bar{p}^2 + m^2}\leq -
\frac{T}{q-1} - \mu\quad {\rm for}\quad q\leq 1; \label{con1}\\
\sqrt{p^2 + m^2}&\geq&- \frac{T}{q-1} + \mu,\qquad \sqrt{\bar{p}^2 + m^2}\geq -
\frac{T}{q-1} - \mu\quad {\rm for}\quad q > 1. \label{con2}
\end{eqnarray}
These lead to  the so-called Tsallis' cut-off prescription for single particle
distributions:
\begin{eqnarray}
n_{q<1}=0 \quad &{\rm for}&\quad \epsilon>\mu+\frac{\beta}{(1\!-\!q)},\quad
\bar{n}_{q<1}=0 \quad{\rm for}\quad {\epsilon}
> - \mu + \frac{\beta}{(1\!-\!q)}\label{lim1a}\\
n_{q>1} = 1\quad &{\rm for}&\quad \epsilon < \mu - \frac{\beta}{(q\!-\!1)},\qquad {\rm
no~limitations~for}\quad \bar{n}_{q>1}. \label{lim1b}
\end{eqnarray}
Notice differences in restrictions imposed on momenta  $p$ and $\bar{p}$ for $q>1$ and
$q < 1$ in Eqs. (\ref{con1}) and (\ref{con2}).
 For $q>1$ and sufficiently small momenta $p$, the limiting energy,
 $\epsilon_{lim}\leq-T/(q-1)+\mu$, gives the usual prescription (\ref{lim1b}) used in \cite{LPQ}.
 For $q<1$ and for large momenta $p$, the limiting energy, $\epsilon_{lim}> -T/(q-1)+\mu$,
  cuts off large momenta,
 well above the Fermi level.
 However, limits given in (\ref{con1}) for anti-fermion momenta practically prevent their
 propagation  (as particle-antiparticle pair excitations) for small temperatures, see Eq.(\ref{lim1a}). Consequently,
 this choice of $S_q$ allow only  for $q>1$ in the propagation of virtual particle-antiparticle pairs.
 This means that for $S_q$ given
 by Eq.(\ref{s1}) only the $q>1$ case has physical meaning in relativistic Fermi gas model.
 Note that relations (\ref{prop}) and (\ref{prop3}) are not now satisfied
 because $e_q(x)*e_q(-x)\neq1$.

In order to allow for a relativistic calculation with $q<1$ and including virtual
particle-antiparticle pairs, we propose to use the entropy functional (cf. \ref{s1}) but
with a different (dual) parameter $q$ for antiparticles, $$q\rightarrow\hat{q}=2-q.$$
This is the second choice to be discussed. In this case we have:
\begin{eqnarray}
\hspace{-8mm}\left(\begin{array}{lll} {\bullet}\\
\bullet\end{array}\right)~~s_{qi}&=&\frac{1-n_{qi}^q-(1-n_{qi})^q}{q-1}+
\left\{n_{qi}\!\rightarrow\!\bar n_{qi},q\!\rightarrow \!\hat{q}\right\}
\hspace{0.4cm}{\rm for}~~0\!<\!n_{qi}(\hat{n}_{qi})<1 \label{s2}
\end{eqnarray}
As a result, after extremalisation, with the same conditions as before, one obtains the
following occupation numbers:
\begin{eqnarray}
n_{q} &=& \frac{1}{e_q\left( x \right) + 1},\qquad \bar{n}_{\hat{q}} =
\frac{1}{\hat{e}_q\left( \bar{x}\right)  + 1}, ~~ {\rm with}~~ \hat{e}_q (x)  = [ 1 +
(\hat{q}-1)]^{\frac{1}{\hat{q}-1}}\label{expq1}.
\end{eqnarray}
Note, that now the corresponding effective occupation numbers are $n^q_q$ for particles
(as before) and $\bar{n}^{\hat{q}}_q$ for antiparticles. The non-extensive analog of
relation (\ref{prop}) is now satisfied because dual $\hat{q}=2-q$ in the anti-fermion
distribution in Eq.(\ref{expq1}) means that $\hat{e}_q(-x)=1/{e}_q(x)$. Therefore
\begin{eqnarray}
n_q(x)+\bar{n}_{\hat{q}}(-x)=1.~~~~~~~\label{prop1}
\end{eqnarray}
To summarize, the $q>1$ relativistic dynamics will be realized in the first choice of
$S_q$, given by Eq.(\ref{s1}), whereas $q<1$ dynamics will be realized in the second
choice of $S_q$, Eq.(\ref{s2}). However, only in the later case is the property
(\ref{prop1}) satisfied.

The third choice of non-extensive entropy consists in removing all previous additional
cut-off`s by using a new definition of entropy, $S_{q}$ (cf. \ref{s1}). It has been
proposed already in \cite{TPM} but without
including anti-fermions. In our case this choice means that:\\
(*) we assume $q>1$ for fermions above the Fermi sea,\\
(*) we assume $q<1$ for fermions in the Fermi sea,\\
(*) we assume $q>1$ for anti-fermions.\\
The resulting $s_{qi}$ has the following form for $q>1$ (cf. \ref{s1}):
\begin{eqnarray} \hspace{-4mm} \left(
\begin{matrix} \bullet\\ \bullet\\
\bullet\end{matrix}\right)~~s_{qi} &=&\left\{
\begin{array}{ll}
\hspace{0.1cm}\frac{1-n_{qi}^{q}-(1-n_{qi})^q}{q-1}+ \left\{n_{qi}\rightarrow \bar
n_{qi}\right\}~~~~~~~~~~~~{\rm
for}~~~0 \leq n_{qi} \leq 1/2 \\
\\ \frac{1-n_{qi}^{\hat{q}}-(1-n_{qi})^{\hat{q}}}{\hat{q}-1}+\left\{n_{qi}\rightarrow
\bar n_{qi},\hat q\rightarrow\!\!q\right\} \hspace{0.8cm}{\rm for}~~~1/2 < n_{qi} \leq
1
\end{array} \right.
\label{s3}
\end{eqnarray}
 Extremalizing $S_q$ with the same conditions as given by Eq.(\ref{qconstraints}) one obtains:
\begin{eqnarray}
n_q&=& \left\{
\begin{array}{lll}
\frac{1}{1+e_q(\beta(\epsilon-\mu))}\hspace{2.5cm}{\rm
for}~~0\leq n_q\leq1/2,\\
\frac{1}{1+e_{\hat{q}}(\beta(\epsilon-\mu))} \hspace{2.5cm}{\rm for}~~1/2<n_q\leq 1,
\end{array} \right.
\nonumber\\
\bar{n}_q &=& \frac{1}{1+e_{{q}}(\beta({\epsilon}+\mu))} \hspace{1.6cm}{\rm
for}~~~0\leq \bar{n}_q\leq1.
\end{eqnarray}
In this case the corresponding effective occupation numbers are $n^{\hat{q}}_q$ for
particles in the Fermi sea, $n^{q}_q$ for particles above the Fermi level
 and $\bar{n}^{q}_q$ for antiparticles.

Because the $q$ parameter is now different (dual) for particles below the Fermi surface
and particles above the Fermi surface, the following non-extensive version of relation
(\ref{prop3}) (obtained similarly to (\ref{prop1})), which connects particle occupations
around the Fermi surface is satisfied:
\begin{eqnarray}
n_q(x)+n_{\hat{q}}(-x)=1~.~~~~~~\label{prop2}
\end{eqnarray}
Note that in a non-extensive Fermi gas one deals with effective occupations,
$n_q^{eff}=(n_q)^q$, as was mentioned before.
 Therefore, relations (\ref{prop1}) and (\ref{prop2}) for the single particle
 distributions  $n_q$ and $\bar{n}_q$
will not be satisfied for the effective distributions $n_q^{eff}$ and $\bar{n}_q^{eff}$.
\footnote{~~~However, the departures from unity are small because
$q\rightarrow\hat{q}=2-q$
 in the power of the effective distribution $n_q^{eff}$.}

To summarize, in this work we derive the fermion $n_q(p)$ and anti-fermion
$\bar{n}_q(\bar{p})$ distributions in the Fermi gas model for 3 different choices of
non-extensive entropy, which correspond to different choices of non-extensive
parameterizations \cite{TPM}. These choices result in different descriptions of dense NM
\cite{SPSA,LPQ,JG,AD1,CW}. In the first two choices, given in Eqs.(\ref{s1}) and
(\ref{s2}), the non-extensivity parameter $q$ is constant for all values of
$n_q(\bar{n}_q)$. Limits imposed by Eq.(\ref{con1}) on the anti-fermion distribution
allows its propagation only for $q>1$. Therefore, in the first choice the particle and
anti-particle distributions can be obtained for the same $q>1$ \cite{LPQ} but without
particle-antiparticle symmetry (\ref{prop1}). This symmetry is restored in the second
choice,  where we choose an anti-fermion distribution with dual $\hat{q}=2-q$ for
anti-fermions  for all values of $n_q(\bar{n}_q)$. The third choice, Eq.(\ref{s3}),
introduces a jump in the $q$ parameter at $x=0$ for particles which results in
discontinuous behavior in the single particle entropy, single particle energy and
effective occupation, $n^{eff}$, which in the non-extensive case has the power $q$
\cite{AD1}.

 The first choice, Eq.(\ref{s1}), realized for $q>1$, with a cut
off (\ref{lim1b}) for small momenta $p$, introduces in general positive correlations
which increase the particle occupation numbers at the bottom of the Fermi sea.
 The second choice, Eq.(\ref{s2}), with  cut-off  (\ref{lim1b}) for large momenta $p$ in the tail of the Fermi
 distribution,
introduces correlations which eliminate large momenta well above the Fermi sea.
However, this scenario is not supported by recent experiments which exhibit strong
short range correlations with a large tail in the distribution above the Fermi sea
\cite{wein}. The third choice, Eq.(\ref{s3}), involves a change of the parameter $q$
in the particle distribution at the Fermi surface. This change of parameter $q$ from
($\hat{q}=2-q<1$) to ($q>1$) means that we expect different, non-extensive effects
below and above the Fermi surface. However, the jump in the $q$ value at the Fermi
surface produces an energy gap near the Fermi surface. Whereas such a gap is well
known in nuclear physics and is produced by strong pairing correlations in the nuclear
ground state below the Fermi surface, it disappears when the temperature increases
\cite{kowal}. Therefore, the above non-extensive description, which assumes a change
of the $q$ parameter at the Fermi surface, acts in the opposite direction and will not
describe this phenomenon. On the other hand, if we modify the change in $q$ in such a
way that it goes smoothly  from $\bar{q}$ in the Fermi sea to ${q}$ above the Fermi
sea, within the finite range of momenta near the Fermi surface, then the occupation
and energy gaps can vanish. In this way we will get the restoration of standard
Boltzman-Gibs statistics, with $q=1$, at the Fermi surface.

Such jumps in $q$ value are absent for antiparticles. One has to stress that their role
is vital in NJL models. In the WM type of approach they are usually neglected because in
these models the effective degrees of freedom are projected onto the positive energy
states (nucleons).  However, the negative energy states (anti-nucleons) cannot be
neglected in NM \cite{SW,Schulze,Ring}. In the relativistic mean field model there are
possible corrections coming from the vacuum polarization which involve contributions from
nucleon-antinucleon pairs \cite{SW,Ring}. Therefore, in the non-extensive version of
these models (like in $q$WM) \cite{Pereira,SPSA,JG}, one has to decide whether higher
order $N\bar{N}$ loops should be included and for $q>1$ (first choice) or ($q<1$) (second
choice). The nuclear EOS described in quark and anti-quark degrees of freedom is
discussed in the $q$-NJL model \cite{JG} approach. Here the propagation of
quark-antiquark pairs, described by $q>1$ statistics, is essential for the dynamics of
massless quarks which change their phase at the critical point. The non-extensive
corrections spread the critical point over a wider area of temperature and density
\cite{JG}.

A description of the EOS in terms of nuclear degrees of freedom is well formulated in
WM \cite{SW}. There are two groups which generalize this mean field approach. The
constant non-extensive parameter $q>1$ is applied in \cite{LP}. The others
\cite{SPSA,JG,CW} apply different distributions which cannot be obtained directly from
extremalization but they satisfy basic thermodynamical relations\cite{CW}. Their $n_q$
distributions are formally identical
 to those obtained from our third choice (and also to those obtained in
 \cite{consistency}) but later they  use  \cite{SPSA,JG,CW} a constant power $q$ in $n_q^{eff}=n_q^q$.
Only in \cite{AD1} the $q$ power changes in the Fermi surface, like in our results for
the third choice given in Eq.(\ref{s3}). Therefore, only the results of \cite{AD1} are
fully consistent.

The following relation for the effective distribution $n_q^{eff}$ shows its behavior
around the Fermi surface - where $n_q(x=0)=1/2$:
\begin{eqnarray}
n_q^{{q}}(x)&+&n_q^q({-x})<1\qquad {\rm for}\qquad q>1 \qquad {\rm 1st~~choice~Eq.(\ref{s1}),} \nonumber\\
n_q^{{q}}(x)&+&n_q^q({-x})>1\qquad {\rm for}\qquad q<1 \qquad {\rm 2nd~choice~Eq.(\ref{s2}),}\\
n_q^{{q}}(x)&+&n_q^q({-x})\simeq1\qquad {\rm for}\qquad q\approx1 \qquad {\rm 3rd~
choice~Eq.(\ref{s3})}. \nonumber \label{prop5}
\end{eqnarray}
Here we assume for the third choice (\ref{s3}) that the jump in parameter $q$ is smeared
for  particle momenta around $x=0$ with $q=1$ at the Fermi surface. These properties
should now be compared with realistic distribution of large nucleon momenta induced by
short range correlations observed recently in \cite{wein}. Such a comparison should allow
for a better understanding of the role of the parameter $q$.

 Concluding, different $q$ parameterizations presented in this work
predict different non-extensive effects in different part of the particle spectra. The
main difference was found  at the Fermi surface between smooth non-extensive dynamics
with constant $q$, (with $q>1$ for the first choice and $q<1$ for the second one),
against the behavior observed in the third choice (\ref{s3}), characterized by a
substantial change in the  non-extensive parameter $q$ across the $q=1$ value on the
Fermi surface. Observations which will indicate the restoration of standard Boltzman-Gibs
statistics ($q=1$) at the Fermi surface would support the third (\ref{s3}) choice which
is used \cite{AD1} in the calculation
 of the EOS with nucleon and meson degrees of freedom. The relations (\ref{prop5})
which reflect the broken symmetries of extensive dynamics, should be accounted for in any
non-extensive analysis in NM.
 It is interesting to note that changes in the non-extensive parameter $q$ (like assumed in (\ref{s3})) can
produce energy gaps, which are in general connected with attractive correlations in Fermi
systems.
\section*{Acknowledgements}
This research  was supported in part by the National Science Center (NCN) under contract
DEC-2013/09/B/ST2/02897. I would like to thank warmly G. Wilk for many fruitful
discussions on the subject. I thank also N. Keeley  for  careful reading of the
manuscript.

\end{document}